\documentstyle[aps,epsfig,floats,twocolumn,amsmath]{revtex}

\begin{document}

 \title{Magneto-polarisability of mesoscopic systems }

\author{Y. Noat $^ {1,2}$,R. Deblock $^1$, B. Reulet $^1$, H. Bouchiat $^1$}

\address{
$^1$ Laboratoire de Physique des Solides, Associ\'e au CNRS, B\^at
510, Universit\'e Paris--Sud, 91405, Orsay, France.\\ $^2$
Kamerlingh Onnes Laboratory, Niels Bohrweg 5, Leiden, Netherland.
 }

\maketitle

\begin{abstract} In order to understand how screening is modified by
electronic interferences in a mesoscopic isolated system, we have
computed both analytically and numerically the average
thermodynamic and time dependent polarisabilities of two
dimensional mesoscopic samples in the presence of an Aharonov-Bohm
flux. Two geometries have been considered: rings and squares.
Mesoscopic correction to screening are taken into account in a
self consistent way, using the response function formalism.

The role of the statistical ensemble (canonical and grand
canonical), disorder and  frequency have been investigated. We
have also computed first order corrections to the polarisability
due to electron-electron interactions.

Our main results concern the diffusive regime. In the canonical
ensemble, there is no flux dependence polarisability when the frequency is smaller than the level spacing. On the other hand, in the grand canonical ensemble  for frequencies larger than the mean
broadening of the energy levels (but still small compared to the level spacing), the polarisability oscillates with flux, with the
periodicity $h/2e$. The order of magnitude of the effect is given
by $\delta \alpha/\alpha \propto (\lambda_s/Wg)$, where $\lambda$
is the Thomas Fermi screening length, $W$ the width of the rings
or the size of the squares and $g$ their average dimensionless
conductance. This magnetopolarisability of Aharonov-Bohm rings has
been recently measured experimentally \cite{PRL_deblock00} and is
in good agreement with our grand canonical result. A preliminary account of this work was given in \cite{pola}.
\end{abstract}

\section{Introduction}

\subsection{Motivation of the paper}

Transport and thermodynamic properties of mesoscopic metallic
samples are known for a long time to be quite sensitive to the
quantum phase coherence of the electronic wave functions at low
temperature \cite{imrybook}. In particular, it was shown that a
mesoscopic ring threaded by a magnetic field exhibits a persistent
current periodic with the flux and with a period of one flux
quantum $\phi_0=h/e$. This current was measured in several
experiments \cite{levy90} \cite{chan91} \cite{mail93},
corresponding to different situations and materials. More
recently, the response of a mesoscopic system to a time dependent
magnetic field was studied both theoretically and experimentally
\cite{reul94} \cite{kame94}.

The purpose of the present paper is to investigate how screening
is influenced by electronic interferences in a phase coherent
sample. In this spirit, we discuss how the electrical
polarisability $\alpha$, i.e. the response of a metallic  sample
to an electrostatic field, a quantity which is directly related to
screening, is also sensitive to mesoscopic effects. $\alpha$ can
be experimentally measured by coupling the mesoscopic samples to a
capacitor. For small voltages, corresponding to the linear regime,
the change in the capacitance is directly proportional to the
average polarisability $<\alpha> $ of the particles.

When the typical size $L$ of the particle is much larger than the
Thomas Fermi screening length $\lambda_s$, $\alpha$ is mostly
determined by the geometrical shape of the particle. For a
spherical particle, it is proportional to the volume $V$ of the
particle, with a small correction of the order of $\lambda_s/L$.

Now, what happens in the case of a quantum coherent sample, (i.e
when $L$ is the order of the phase coherence length)? Is there a
{\em mesoscopic correction} to the polarisability, due to
contribution from electronic interferences?

In order to answer this question, we have calculated the flux
dependent correction to the average polarisability, which we call
``mesoscopic correction'', for {\em two dimensional mesoscopic
samples} in the presence of an {\em Aharonov-Bohm flux}. Two
different geometries have been considered, namely a two
dimensional ring or square in an in-plane electric field. The role
of statistical ensemble, disorder, and finite frequency have been
investigated. In the case of a time dependent electric field, we
distinguish two different limits, whether the frequency is smaller
or larger than the inverse relaxation time, or mean level
broadening $\gamma $. The case $ \omega \ll \gamma $ corresponds
to the thermodynamic limit whereas $\omega \gg \gamma $
corresponds to the finite frequency limit.

\subsection{Scope of the paper}

This paper is organized as follows: the definition of the
electrical polarisability and the classical result for a
macroscopic sample are given in section \ref{basics}.

In section \ref{pol_no_screening}, we derive the expression of the
electrical polarisability of a system of non-interacting
electrons, in the spirit of the early work of Gorkov and
Eliashberg (GE) \cite{gorko65}. A giant unphysical polarisability is
obtained. However it is possible to discuss in this very simple
model the flux dependence of two quantities which contribute to
the polarisability, namely the matrix element of the position (or
screened potential) operator and the contribution of energy
levels. Then, we examine in section \ref{screening} how screening
can be taken into account. In particular, we show how to
incorporate contributions of electronic interferences. Using those
results, a general expression for the polarisability and its
mesoscopic correction for a time varying electric field  are
derived in section \ref{expressions}.

 We study  then separately the case of the grand canonical
and canonical statistical  ensemble. The grand canonical average
polarisability is found to depend only on the flux dependent
matrix elements of the screened potential whereas the Canonical
average depends also on the energy levels. As a result  when the frequency is smaller than the level spacing the  magneto-polarisability is found  much smaller in the canonical ensemble compared 
	to the  grand canonical case. Finally, we present results on the
Hartree-Fock correction to the zero frequency polarisability.
Section VI is devoted to a comparison of these results with recent
experiments.

\section{Basic concepts}
\label{basics}

\subsection{Screening and polarisability}
\label{screepol}

When an external field  ${\bf{ E_{ext}}}=-\bf{grad}(\phi_{ext})$
is  applied on a conductor, the charge density adjusts in such a
way that the effective field inside the conductor cancels. This
phenomenon is called screening. In principle the computation of
the potential inside the sample requires to solve a complex N body
hamiltonian taking into account  the long range Coulomb
interaction between  electrons. However it is possible in general
to treat the interactions in a mean field approximation. The
screened potential inside the conductor is then  given by:

\begin{equation}
\phi({\bf r_1})=\phi_{ext}({\bf r_1})+\int \delta \rho({\bf
r_2})U({\bf r_1},{\bf r_2})dr_2 \label{relself}
\end{equation}
Where $\delta \rho({\bf r_2})$ is the charge density induced at
the point $\bf r_2$ and $U({\bf r_1},{\bf r_2})$ is the coulomb
interaction potential. In the linear response limit, $\delta
\rho({\bf r_2})$ can be described by the electric response
function: $\chi({\bf r_1},{\bf r_2})$, which relates the variation
of charge in the system at $\bf r_2$ to a local  perturbation at
$\bf r_1$:$\delta({\bf r_1})$.

\begin{equation}
\delta \rho({\bf r_2})= \int \chi({\bf r_1},{\bf r_2}) \phi({\bf
r_1})dr_1
\end{equation}

Relation \ref{relself} can then be written more conveniently in a
matrix form:

\begin{equation}
\phi=(1-U\chi)^{-1} \phi_{ext} \label{relmat}
\end{equation}
In a clean  infinite system, the induced charge density is simply
related to the screened potential $\phi$ in Fourier space: $\delta
\rho(q)=\chi(q)\phi(q)$, where $\phi(q)=\int e^{i{\bf q}.{\bf
r}}\phi({\bf r})dr$. The problem is more complex in a disordered
system, which does not have the translational invariance. In any
case  for a finite system, it is possible to compute the response
function from the eigenstates $|\alpha>$ of the system in the
absence of the external field, associated to the eigenvalues
$\epsilon_\alpha$:
\begin{equation}
    \chi({\bf r_1},{\bf r_2})=\sum_{\alpha = 1}^N \sum_{\beta < \alpha } 		\frac{\psi_{\alpha}^{*}({\bf r_1}) \psi_{\beta}({\bf r_2}) \psi_{\alpha}({\bf r_1}) 
\psi_{\beta}^{*}({\bf r_2})}{\epsilon_\beta-\epsilon_\alpha}+ c.c.
\end{equation}
In most cases we will identify these eigenstates as the solution
of the hamiltonian of the system  without electron interactions
which will be taken into account only by considering the screening
on the applied field. The effect of interactions on the
eigenstates  will be however discussed in section VII within the
Hartree Fock approximation.

The application of an electric field  on a finite metallic system,
results in an induced dipolar moment ${\bf P}$ which, for a
sufficiently small field \footnote{This point will be discussed
later on}, is proportional to the applied field.

\begin{equation}
{\bf P}=\int \rho \left({\bf r} \right) {\bf r} d{\bf r} =\alpha
{\bf E} \label{defmdip}
\end{equation}
where $\alpha$ is by definition the polarisability of the
conductor, which can be expressed as a function of the response
function:

\begin{equation}
\alpha=\frac{1}{E^2}\int \chi({\bf r_1},{\bf r_2}){\bf
E.r_2}\phi({\bf r_1})dr_1dr_2
\end{equation}

\subsection{Classical polarisability}
\label{polclass}
\subsubsection{  Plate with  perpendicular electric field }
The polarisability of a macroscopic piece of metal is described by
classical electrostatics. In this approximation, the dipolar
moment can be calculated by solving Poisson equation, $\Delta
V+\frac{\rho}{\epsilon_0}=0$, in all space and imposing as well
that the conductor is at a constant potential. The problem is
solvable analytically for a sphere of volume $V$: $\alpha=\epsilon_0
(4\pi/3) V $. It is also possible to estimate the polarisability
of a metallic plate of thickness $a$ in a constant electric field
perpendicular to the plate. The induced charge density is such
that the electric field cancels inside the plate. Neglecting
border effects, we have:
\begin{equation}
\rho \left( {\bf r}\right)=\frac{E \epsilon_0}{2}
\left[\delta\left(x-\frac{a}{2}\right)-\delta\left(x+\frac{a}{2}\right)
\right]
\end{equation}

Integrating equation \ref{defmdip}, one finds the classical
Polarisability:
\begin{equation}
\alpha=\epsilon_0 a^3=\epsilon_0 V
\end{equation}

\subsubsection{2d and 1d systems in longitudinal electric fields}

We now examine the  case of 2d conductors, and more specifically a
strip of length $L$ and width $W$, with $L\gg W$, in an in-plane
electric field perpendicular to its long axis. In order to
determine the induced charge density, the effective electric field
inside the strip is imposed to be zero. In the limit of an
infinite strip, this condition leads to the following equation:

\begin{equation}
\int \frac{1}{2\pi\epsilon_0}\frac{\rho (x')}{x-x'}dx'-E=0
\label{equastrip}
\end{equation}

\begin{figure}
\centerline{ \epsfxsize 7cm \epsffile{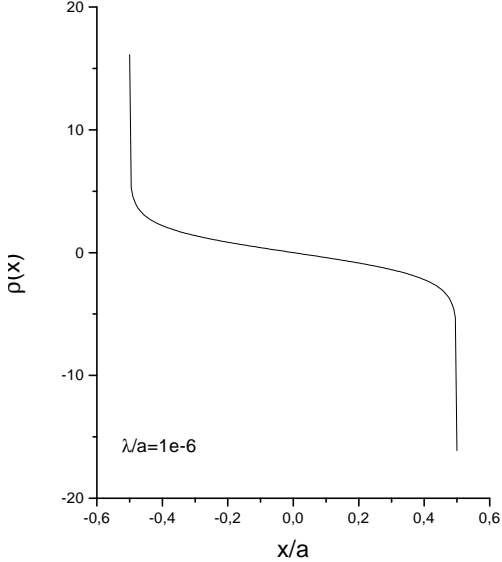}}
\caption{Induced charge density of a conducting strip in an
in-plane uniform electric field: the function diverges
logarythmically as a function of the cutoff $\lambda/a$.}
\label{fruban}
\end{figure}

We have calculated numerically the solution of \ref{equastrip} in
the case of an infinite strip (fig. \ref{fruban}).
In contrast with the 3 dimensional case, the charge distribution
spreads over the full width of the strip in order to cancel the
external field. The numerical solution is very close to a
logarithm. Assuming that the density is of the form
$\rho(x)\approx EW \log\left(\frac{W}{2}-x\right) /
\log\left(\frac{W}{2}+x\right)$, the integral \ref{defmdip} of the
previous function gives $\alpha \approx \epsilon_0 W^2 L$.

This result extrapolates to a 2 dimensional sample of typical size
$W$, for which the in-plane polarisability scales like $\alpha
\sim \epsilon_0 W^3$. For instance the polarisability of a disk of
radius $R$ is $\alpha=\epsilon_0\frac{16}{3}R^3$ \cite{landau}.

In the case of a one dimensional wire, whose length L is much
longer than its diameter D, in an electric field along its axis,
the polarisability  has been shown to be $\epsilon_0 L^3 ln(D/L)$\cite{landau}.
Similarly as in the $2d$ case, the charge density has to spread
over the entire wire in order to screen the external potential.

The case of a 1D ring is also  exactly solvable. In order to take
screening into account, we first assume that the induced charge
density in the ring is of the form $\rho(\theta)=\lambda
\cos(\theta)\delta(z)\delta(\rho-R)$, from which we deduce the
induced potential in the ring:

\begin{equation}
\phi_{ind}(\theta)=\int_{0}^{2\pi}
\frac{1}{8\pi\epsilon_0}\frac{\rho(\theta')}{\left|\sin\left(\frac{\theta-\theta'}{2}\right)\right|}d\theta'
\end{equation}

The integral is logarythmically diverging and it is necessary to
introduce a cutoff $\theta_c=W/R$, related to the finite width W of the
ring: $\phi_{ind}(\theta)\approx -\frac{\lambda
\cos(\theta)\ln\left(\theta_c\right)}{4\pi\epsilon_0}$.

The screened potential reads $\phi(\theta)=\cos(\theta)\left(E+\lambda J\right)$, 
with $J=\frac{1}{4\pi\epsilon_0}\ln\left(\frac{1}{\theta_c}\right)$.
The polarisability is obtained either from the potential
calculated above, either from the expression of the charge. This
self-consistent relation sets the value of the parameter
$\lambda$, from which we deduce the polarisability of the ring.

\begin{equation}
\alpha=\frac{\alpha_0}{1+\frac{2J\alpha_0}{R^3}}
\end{equation}

where $\alpha_0=\epsilon_0 \pi^2 R^3$ which reads  in the limit
$J\gg 1$:

\begin{equation}
\alpha\approx\frac{\epsilon_0 \pi^2
R^3}{\ln\left(\frac{R}{W}\right)}
\end{equation}

As a result both longitudinal polarisabilities  of a ring and a
disk scale like the cube of their radius, just like the
polarisability of a sphere.

\subsection{A  mesoscopic correction to the polarisability ?}

In the following, we define the quantity we wish to calculate. As
we have just seen, the classical polarisability of a metallic
sample is determined by geometrical factors. However, due to the
Pauli principle, charges can not strictly accumulate on the border
of the sample, but the charge density distribution rather extends
on the screening length $\lambda_s$. This effect reduces the
polarisability compared to its classical value by a quantity  $\delta\alpha_{TF}$  of the order of
$\lambda_s/L$, where $L$ is the typical size of the conductor
along the electric field. In a coherent sample, electronic
interferences might also give rise to a contribution $\delta
\alpha_Q $ to the polarisability. The total polarisability can
thus be written: $\alpha=\alpha
_{cl}+\delta\alpha_{TF}+\delta\alpha_Q$. In order to show evidence
of this correction, a magnetic flux can be applied through the
sample in order to introduce a flux dependent part in the phase
difference of electronic trajectories.  We will call ``mesoscopic
correction'' the flux dependent part of the polarisability $\delta
\alpha (\Phi)$. We will focus on the ensemble average of this
quantity $\overline{\delta \alpha} (\Phi)$, which is $\Phi_0/2$
periodic. We will characterize the flux dependence of
$\overline{\delta \alpha}$ by $\delta_\Phi (\alpha) =
\overline{\delta \alpha} (\Phi_0/4) - \overline{\delta \alpha}
(0)$.

\subsection{Model used for the numerical simulations}

Part of the analysis we have done rely on computer simulations.
The two dimensional rings or squares are modelled by the Anderson
model. Each atomic site is coupled to its nearest neighbour by a
hopping term $t$. Disorder is introduced by on-site energies
randomly distributed in the interval $[-w,w]$. The Anderson
hamiltonian can be written in second quantization:

\begin{equation}
H=\sum_{k}^{}\epsilon_kc^\dagger_kc_k+te^{i\phi_k}c^\dagger_kc_{k+1}+
te^{-i\phi_k}c^\dagger_{k+1}c_k
\end{equation}

where $c^{\dagger}_k$ is the creation operator associated with
site $k$. The magnetic flux is introduced in this hamiltonian through
the phase factor of the hoping element with $\phi_k=2 \pi \Phi/\Phi_0 \, 
(x_k-x_{k+1}) / L$ for a ring of size $L$. In the case of squares  we have taken into account the penetration of the magnetic field in the sample: the phase factor appearing in the hopping matrix elements  is computed from the integral of the potential vector between concerned neighbors. In two dimensions, 
an estimation of the elastic mean free path is given by 
$l_e\approx 30\left(\frac{t}{w}\right)^2$ \cite{reul94} 
allowing us to choose values of $t/w$ so that the
system is in the diffusive regime. In order to average quantities we take 
advantage of ergodicity properties in the diffusive regime and average over the 
number of electrons in the system between one quarter  and three quarter 
filling.

\section{Quantum polarisability of a system of non-interacting electrons}
\label{pol_no_screening} We now examine the electrical response of
a quantum system of non-interacting electrons, which was first
calculated by GE \cite{gorko65} for a
metallic grain both in the diffusive and ballistic limit. Their
conclusion is quite surprising: in the diffusive regime, $\alpha$
is bigger than the classical polarisability by a factor
$(a/a_0)^2$, where $a$ is the typical size of the grain and
$a_0=h^2/(m e^2)$ is the Bohr radius. Later on, Rice
\cite{rice73} stressed the lack of screening in the calculation of
GE and showed that when screening is taken into account, one
recovers the classical polarisability. We will see in the
following that it is useful to consider the flux dependence of the
polarisability  within this crude approximation, since many
results remain qualitatively true in the presence of screening.

\subsection{Position of the problem}

 The
polarisability of a system of non-interacting electrons can be
understood as the sensitivity of the energy spectrum to an
external electric field, just like persistent currents measure the
sensitivity of the spectrum to an Aharonov-Bohm flux . As a matter
of fact, the eigenstates $|\alpha>$ of the system  and  the
eigenvalues $\epsilon_\alpha$, are modified by an external
electric field $E$.
\begin{equation}
\epsilon_{\alpha}'=\epsilon_{\alpha}-<\alpha|e{\bf
E.r}|\alpha>+\sum_{\beta \neq \alpha}\frac{{\left|<\alpha|e{\bf
E.r}|\beta>\right|}^2}{\epsilon_\beta-\epsilon_\alpha}+ ...
\label{eqener}
\end{equation}

\[
|\alpha>' \approx |\alpha>+\sum_{\beta \neq
\alpha}\frac{<\beta|{e\bf
E.r}|\alpha>}{\epsilon_\beta-\epsilon_\alpha}|\beta>+...
\]

Consequently, it exists an induced charge density $\delta
\rho({\bf r})=e \sum_{\alpha=1}^N {\left|\psi'_\alpha({\bf
r})\right|}^2-{\left|\psi_\alpha({\bf r})\right|}^2 $, associated
with the asymmetry of each wave function generated by $E$. Using
the definition \ref{defmdip} of $\alpha$, the expression of the
polarisability of a system of non-interacting electrons at zero
temperature can be found:

\begin{equation}
\alpha = \frac{2e^2}{E^2}\sum_{\alpha=1}^{N}\sum_{\beta \neq
\alpha}\frac{{\left|<\alpha|{\bf
E.r}|\beta>\right|}^2}{\epsilon_\beta-\epsilon_\alpha}
\label{polsansint}
\end{equation}
This expression  depends on the eigen-energies of the unperturbed
system and its eigen-functions through the matrix elements of the
position operator. Due to the energy denominator, the
polarisability will in particular be  very sensitive to the
electric field induced coupling between the last occupied levels
and  the first non occupied ones. Note also that this expression
of the polarisability is also  from expression \ref{eqener}
identical to the second derivative of the total energy of the
system with respect to  electric field.
\[ \alpha = -\sum_{\alpha=1}^{N}\frac {\partial^2 \epsilon_\alpha}{\partial E^2}
\]
These results  present strong similarities with  the paramagnetic
contribution of orbital susceptibility of a mesoscopic ring pierced 
by a flux line which is identical to expression \ref{polsansint} where the ${\bf
E.r}$ is replaced by  ${e\bf A.p}$ where ${\bf A}$ is the
potential vector and ${\bf P}$ is the kinetic momentum operator.
The total magnetic  susceptibility contains also a constant
diamagnetic term $ -Ne^2/m$ \cite{trive88},\cite{reul94} which does not
exist in the electric response.

\subsection {1 dimensional Aharonov-Bohm ring}
Using expression \ref{polsansint} it is possible to calculate
exactly the induced dipolar moment in a 1d non-disordered ring in
an in plane electric field .

Wavefunctions only depend on the angle $\theta$ ,indicating the
position in the ring, and satisfy the 1d Shr\"odinger equation:
$-\frac{\hbar^2}{2mL^2}\frac{\partial^2}{\partial\theta^2}\psi(\theta)=E\psi(\theta)
$. Furthermore, the magnetic flux, associated with the periodicity
in the ring, imposes the following boundary condition $
\psi(\theta+2\pi)=e^{i2\pi\frac{\phi}{\phi_0}}\psi(\theta) $. This
equation can be solved. Using the parity and periodicity of the
spectrum, we then order the eigenvalues and the corresponding
wavefunctions in ascending order in the interval
$\left[-\frac{\phi_0}{2},\frac{\phi_0}{2}\right]$:

\begin{equation}
\left\lbrace
\begin{array}{lll}
\epsilon_{2p}=\frac{\hbar^2}{2mL^2}{\left(p+\frac{\phi}{\phi_0}\right)}^2
& \psi_{2p}(\theta)=\frac{1}{\sqrt
L}e^{i\theta\left(p+\frac{\phi}{\phi_0}\right)} \\
\epsilon_{2p+1}=\frac{\hbar^2}{2mL^2}{\left(p-\frac{\phi}{\phi_0}\right)}^2
& \psi_{2p+1}(\theta)=\frac{1}{\sqrt
L}e^{i\theta\left(-p+\frac{\phi}{\phi_0}\right)} \\
\end{array}
\right.
\end{equation}

As a consequence of the particular geometry of the system, the
electric field only couples certain states. In particular, the
matrix element between adjacent states cancels unless the quantum 
numbers $p$ et $q$ are such that $p-q=\pm 2$, in this case : 
$<p|X|q>=\frac{R}{4 \pi}$ 
The polarisability depends on the parity of the number
of electrons $N$ and reads :

\begin{multline}
\alpha_N(\Phi) = \frac{e^2 R^2}{8 \pi^2} \left( 
\frac{1}{\epsilon_{N+2}(\Phi)-\epsilon_{N}(\Phi)} \right.\\
\left. -\frac{1}{\epsilon_{N+1}(\Phi)-\epsilon_{N-1}(\Phi)} \right)
\end{multline}

At zero flux and for $N \gg 1$: $\alpha_N=\frac{1}{8\pi^4}\frac{L^4}{a_0}\frac{1}{N}$.  
It is possible to evaluate $\delta_{\Phi} \alpha$ by computing 
$(\alpha_{N+1}-\alpha_{N})/2$. As a result :

\begin{equation}
\frac{\delta_{\Phi} \alpha}{\alpha} \approx -\frac{1}{N^2}. 
\label{dpol1d}
\end{equation}

It is a very small effect, since it decreases rapidly with the
number of electrons in the ring. This result remains  true for a
multichannel ring. We attribute this effect to selection rules
inherent to the square lattice which  are responsible for the
cancellation of the matrix elements of operator X between
eigenstates close to a level crossing. In particular it has been
shown \cite{fulde} that these selection rules do not exist in the
hexagonal lattice where giant magnetopolarisability is expected  for particular values of flux at
the same level of approximation.

\begin{figure}
\centerline{ \epsfxsize 6cm \epsffile{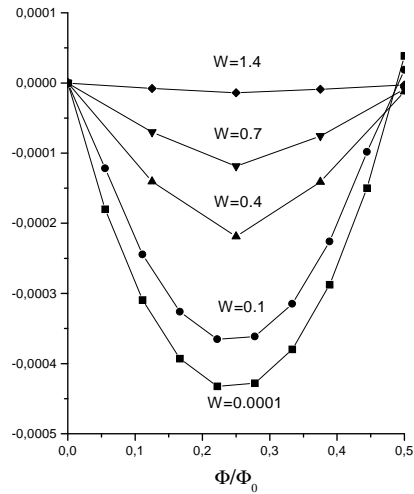}}
\caption{Magnetopolarisability of a 1d ring, of length $L=100$,
calculated with the Anderson model, for several values of the
disorder: $w=10^{-4}$; $0.1$; $0.4$; $0.7$; $1.4$.} \label{p100}
\end{figure}

In order to investigate the effect of disorder we have performed
numerical simulations using the Anderson model. On figure
\ref{p100} is plotted the magnetopolarisability for a ring of
length $L=100$ and several values of the disorder $w$. In absence
of disorder, the results are in qualitative agreement with formula
\ref{dpol1d}. The effect then decreases when the disorder is
increased, with a law close to $\frac{1}{w^2}$. One is tempted to
attribute this result to occurrence of localization in the 1D ring.
However we will see in the following that this cancellation of the flux
dependent polarisability is also observed in a multichannel ring
or disk in the diffusive regime.

\subsection {Diffusive system}
\begin{figure}
\centerline{ \epsfxsize 6cm \epsffile{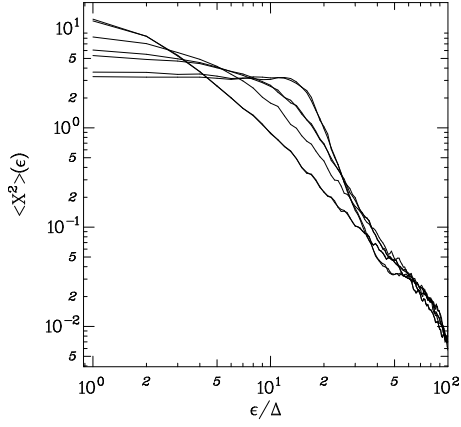}}
\caption{Energy dependence of the non diagonal matrix elements in
a ring of length $L=60$ with $M=4$ channels, and several values of
the disorder:$w=0.9$; $1.2$; $1.4$; $1.7$. Note the additional
step  observed for $\epsilon\approx 5\Delta$, which corresponds to
the Thouless energy in the transverse direction of the ring:
$\frac{hD}{W^2}$where  $W$  is the width of the ring.}
\label{xener}
\end{figure}

In the following we discuss a diffusive system characterized by a
diffusion coefficient $D=(1/d) v_F l_e$ where $l_e$ is the elastic
mean free path which is assumed to be shorter than the system
size $a$ along the electric field. It has been shown \cite{gorko65}, \cite{entin90}, \cite{frahm90} using semiclassical arguments that
the average square   matrix element,
${\left|<\alpha|X|\beta>\right|}^2$ depends  mainly  of the energy
difference $ \epsilon=|\epsilon_\beta-\epsilon_\alpha|$ 
and the Thouless energy $E_c=h D/L^2$. For
$\epsilon < E_c$ it is  of the order of $a/g$ where $g=E_c/\Delta$ is the ratio
between the Thouless energy  and the mean level spacing $\Delta$.
At higher energy  it decreases like $1/\epsilon^2$.  This behavior
is illustrated in fig.\ref{xener}  showing numerical results on a
sample for different values of disorder where these  two regimes
can be clearly distinguished.

It is then easy to deduce the order of magnitude of the
polarisability noting that the summation in expression
\ref{polsansint}  can be  restricted to $|\alpha-\beta|<g$
as a result:

\begin{equation}
\alpha \sim e^2 a^2/\Delta
\end{equation}

The GE result is  then recovered assuming that
the Fermi wave length is of the order of the Bohr radius.
We will see  in the next section  that this  result
is modified by screening. It is however  worth continuing this
analysis in this oversimplified picture of non interacting
electrons in the case of an Aharonov-Bohm ring.   From expression
\ref{polsansint} the polarisability is  expected to exhibit  flux
dependence both from the matrix elements and energy denominators.
We discuss in the following separately these two contributions
which is justified in the context of random matrix theory where
eigenfunctions and eigenenergies constitute two sets of
independent random variables \cite{metha}.

\subsubsection{Energy denominator}

We focus on the flux dependence of the quantity $K=\left<\sum_{
\alpha <\beta}^{}\frac{1}{\epsilon_\beta-\epsilon_\alpha}\right>$. This
quantity depends only on the energy differences
$(\epsilon_\alpha-\epsilon_\beta)$, and can be expressed as a
function of $R(\epsilon)$, the two levels correlation function :
\begin{equation}
K=\int_0^{E_{max}}d\epsilon\int_0^\epsilon du \frac{R(u)}{u}
\end{equation}

\begin{figure}
\epsfxsize 8cm \epsffile{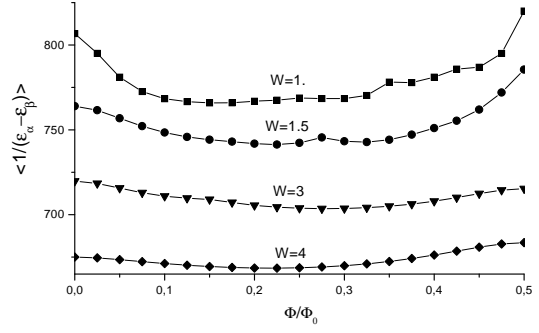} \caption{K calculated
numerically in a ring $80\times 8$, for different values of the
disorder: $w=1.$; $1.5$; $3$; $4$.} \label{denom}
\end{figure}

In the diffusive regime, $R(s)$ is well described by random matrix
theory \cite{metha} and the average probability to find two
degenerate adjacent levels is zero. This property is
characteristic of level repulsion in the spectrum of a random
matrix, which is stronger in a system where reversal symmetry is
broken compared to a system where it is not.  As a consequence,
$K$ decreases as a function of the flux at low magnetic flux and
increases back in the vicinity of $\phi_0/2$ where time reversal
symmetry is recovered.

We have calculated numerically $K$ in a disordered ring of length
$L=80$ and width $a=8$ for several value of the on-site disorder
$w$ corresponding to the diffusive regime. Figure \ref{denom}
illustrate the effect described above. In the absence of other
contributions, this term would give rise to a negative
magnetopolarisability.

\subsubsection{Matrix element}

\begin{figure}
\epsfxsize 6.cm \epsffile{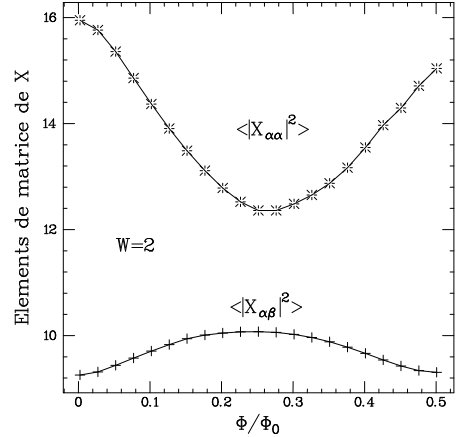} \caption{Flux
dependence of the diagonal and non-diagonal matrix elements in a
ring of length $L=60$ and $M=4$ channels, for $w=2$. The factor
$2$ at zero flux between the diagonal and non-diagonal matrix
elements is due to the fact that only the states such that $
\epsilon_\alpha < \epsilon_\beta $.} 
\label{xdnd}
\end{figure}

On figure \ref{xdnd} are plotted the  flux dependance of the
average square of the diagonal and non-diagonal matrix elements
for a  diffusive ring. It can be noted that they have 
opposite flux dependences, resulting from the fact that
$Tr\left(X^2\right)=\sum_{\alpha}X_{\alpha\alpha}^2+\sum_{\alpha\neq\beta}X_{\alpha\beta}^2$
is flux independent. The diagonal elements decrease as a
function of the flux, whereas the non-diagonal elements increases.

\begin{figure}
\centerline{ \epsfxsize 7cm \epsffile{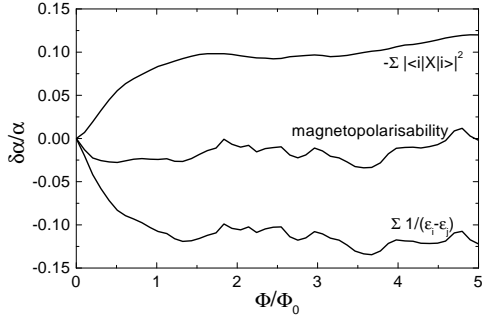}}
\caption{Flux dependence  of the matrix element and energy
denominators contributions  to  the average unscreened
polarisability  of a disordered square $30\times 30$, calculated
with the Anderson model for $w=3$. These 2 contributions nearly
compensate yielding to the absence of magnetopolarisability. } \label{polacgc}
\end{figure}

The time reversal property of the operator ${\bf X}$ implies  that
its diagonal matrix elements are real even function of flux and
can be developed in successive powers of $cos(2\pi \phi/\phi_0)$.
$<X_{\alpha,\alpha}^2>$ is maximum for multiple values of
$\phi_0/2$. It is  possible to evaluate analytically  this flux
dependence of the diagonal matrix elements  from random matrix
theory, using  the relation between  $X_{\alpha,\alpha}$ and the
sensitivity of the energies to an electric field:
$X_{\alpha,\alpha}= \left( \frac{\partial\epsilon_\alpha}{\partial E} \right)_{E=0}$.
Since  electric field preserves time-reversal symmetry, the
typical value of the derivative of the energy levels with respect
to the electric field
$<{\left|\frac{\partial\epsilon_\alpha}{\partial E_0}\right|}^2>$
is proportional to $\frac{1}{\beta}$. Consequently: $
{\left|X_{\alpha,\alpha}\right|}^2(\phi=\frac{\phi_0}{4})=\frac{1}{2}{\left|X_{\alpha,\alpha}\right|}^2\left(\phi=0\right)
\label{xdiag} $. Figure \ref{xdnd} shows that this last result is
 qualitatively true.
Therefore, in absence of the energy denominator
$\frac{1}{\epsilon_\beta-\epsilon_\alpha}$, the flux dependence of
the matrix elements  would give rise to a positive
magnetopolarisability. As a result this flux dependence is
opposite to the contribution of the energy denominators. Moreover
we can see on fig. \ref{polacgc} that these contributions almost
exactly cancel and there is no magneto-polarisability in this
model.

We will see however in section V that a non zero
magnetopolarisability is found at finite frequency in the grand
canonical Ensemble or at finite temperature in the canonical
Ensemble where the energy denominators contribution disappears.
At this stage it is  interesting to emphasize  that the  flux
dependence  of the average square of the X operator matrix element 
is opposite to the same quantity related to the current operator which 
changes sign by time reversal symmetry. So when computing the magnetic
susceptibility  the contribution of the matrix elements and energy
denominators are of the same sign. This effect is related to  the
existence of a finite average current in the canonical ensemble for a
diffusive ring.

Note that both persistent current and zero frequency
polarisability can be expressed as a function of the free energy
of the system: $I=-\partial F(E,\phi)/\partial \phi$ and
$\alpha=-{\partial^2 F(E,\phi)}/{\partial E^2}$. So:
\begin{equation}
\frac{\partial \alpha}{\partial
\phi}=\frac{\partial^2I_{per}}{\partial E^2}
\label{iperalpha}
\end{equation}

Let us emphasize that the  absence of  thermodynamic magnetopolarisability    thus related to the insensitivity of persistent currents to an electrostatic field and can be qualitatively
understood using a semi-classical argument. The effect of the electrostatic potential
slowly varying at the scale of the Fermi wavelength can be
included into  the phase of the electronic wavefunctions.  Persistent currents can be computed from the   integral of the classical action on a circular  orbit. The   contribution of the static  electrostatic potential on such an orbit: $(e/\hbar v_f)\int V({\bf r})ds$ where  $ V({\bf r})= E r cos(\theta)$ and  $ds=r d\theta$ obviously cancels out.  This argument is helpful 
to understand why the polarisability is independent of the flux in this 
very low frequency regime.

\section{Screening}
\label{screening}

\subsection{The Thomas-Fermi approximation}

As mentioned above the mean field approximation reduces the
complicated  many body problem of interacting electrons to a much
simpler one, in which electrons move in an effective potential
$\phi({\bf r})$, resulting from the screening of the applied
potential  by other electrons. Therefore, it goes beyond a
simple electrostatic calculation by taking into account the
kinetic energy of electrons as well as their fermionic character \cite{ash-mer}.

In the linear regime, i.e for a small external potential, the
induced charge density is simply proportional to the effective
potential \cite{berko92}, \cite{berk92}:
\begin{equation*}
\delta \rho({\bf r})=-e^2 n(E_F)\phi({\bf r})=-\epsilon_0
k_s^2\phi({\bf r})
\end{equation*}
 at 3D with $k_s=\sqrt{\frac{e^2n(E_F)}{\epsilon_0}}$ and
\begin{equation*}
\delta \rho({\bf r})=-e^2 n(E_F)\phi({\bf r})=-\epsilon_0
k_s\phi({\bf r})\delta (z)
\end{equation*}
for a conducting plane at z=0 with $k_s=\frac{e^2n(E_F)}{\epsilon_0}$. 

The resolution of the self-consistent equation \ref{relself}
within this Thomas Fermi approximation for a 3D sample
in the presence of a uniform external applied field, gives rise to a screened
potential whose value is  of the order of $E/k_s$ which is
confined to the border of the sample within $\lambda_s=1/k_s$. This
result is in principle not true any more in a 2D system where
classical screening already involves charge displacement in the
whole system. However since the charge distribution is always
singular on the edge of the sample it is  possible to approximate
the screened potential for a disk of radius R by: $\Phi(r,\theta)=
(E/k_s) F( (R- r))cos(\theta)$ where F is a peaked function
centered on zero of width $1/k_s$.
When screening is taken into account the expression of the quantum
polarisability at T=0 reads:
\begin{equation}
\alpha = \frac{2e^2}{E^2}\sum_{\alpha\neq\beta}^{} \frac{<\beta
|\phi_{T.F}({\bf r})| \alpha><\beta|{\bf
E.r}|\alpha>}{\epsilon_\alpha-\epsilon_\beta}
\end{equation}
Within this approximation the GE result for the
polarisability is modified by a factor $1/(a k_s)^2$ and becomes
identical to the classical result.

\subsection{Beyond The Thomas-Fermi approximation}

\label{beyondtf}

As we have seen, the Thomas-Fermi approximation does not take into
account contribution of the screening due to electronic
interferences. They indeed  give rise to a quantum correction for
the response function $\chi$, and consequently to the effective
potential $\phi$ and the polarisability $\alpha$:

\[
\left\lbrace
\begin{array}{l}
\chi=\chi_{T.F}+\delta \chi \\

\phi=\phi_{T.F}+\delta \phi \\

\alpha=\alpha_{T.F}+\delta \alpha
\end{array}
\right.
\]

where $\chi$ is the response to the local field. Assuming that
those corrections are small compared to the Thomas-Fermi value, it
is possible to show, in agreement with Efetov \cite{efe96} (see
appendix \ref{calculefetov} for the detailed calculation), that
$\delta \alpha$ can be simply expressed as a function of the
mesoscopic correction to the one electron response function
$\delta \chi$ and the Thomas-Fermi potential:
\begin{equation}
	\delta \alpha \approx \frac{1}{E^2}Tr \left( \phi_{T.F}\delta \chi \phi_{T.F} \right)
\end{equation}
In the following  we discuss the response to a time dependent
electric field. We will see that dynamical polarisability can be
very different from  the static one.
\section{Expression of the polarisability}
\label{expressions}

\subsection{Response to  time-dependent potential}

The application of a time-dependent external potential $V(t)= e{\bf
E.r}\exp\left(i\omega t\right)$ raises the problem of the relaxation of the system towards equilibrium. This process is made by inelastic processes
characterized by a typical time scale, related to 
inelastic collisions. 

In the limit of a weak coupling, this process can be described by
a master equation on the density matrix \cite{trive88}:

\begin{equation}
i\hbar \frac{\partial\rho}{\partial
t}=[H_0+V(t),\rho]-i\gamma\left(\rho-\rho_{eq}\right)
\end{equation}

The parameter $\gamma$ represents the broadening of the energy
levels and characterizes dissipation processes, allowing the
relaxation of the system towards equilibrium. The density matrix $\rho_{eq}$ satisfies
the condition $\left[H,\rho_{eq}\right]=0$, $H=H_0+e{\bf
E.r}\exp\left(i\omega t\right)+\phi_{ind}({\bf r},\gamma,t)$ being
the hamiltonian of the system in the mean field approximation, in
an AC electric field at frequency $\omega$. Two limits are to be considered.
When the frequency of the time dependent potential is small compared to $\gamma$; the system follows
the potential and stays at every moment at equilibrium. On the
contrary, at high frequency compared to $\gamma$ , the system is always out of
equilibrium.

In addition, we assume that the effective potential is always in
phase with the external one. This is justified if the dissipative
part of the polarisability is very small compared to the
non-dissipative part ($\alpha''\ll \alpha')$, this last quantity
being discussed in \cite{nrbm98}.

The expression of the density matrix is then obtained by solving
the master equation. From it, we deduce the response function and
later on the polarisability and its quantum correction $\delta
\alpha_{\phi}$, which can be expressed as a function of the eigenstates
and eigenvalues of the hamiltonian $H_0$.

\begin{multline}
\alpha = -\frac{2e^2}{E^2}Re \Bigg( \\
 \sum_{\alpha\neq\beta}
\frac{f_\alpha-f_\beta}{\epsilon_\alpha-\epsilon_\beta}
\frac{\epsilon_\alpha-\epsilon_\beta-i\gamma}{\epsilon_\alpha-\epsilon_\beta+ \omega-i\gamma}
<\beta |\phi_{T.F}| \alpha><\beta|{\bf E.r}|\alpha> \\
+\frac{\gamma}{\gamma+i\omega}\sum_{\alpha}^{}\frac{\partial
f_\alpha}{\partial\epsilon_\alpha}<\alpha|\phi_{T.F}|\alpha>
<\alpha|{\bf E.r}|\alpha> \Bigg)
\label{pol}
\end{multline}

\begin{multline}
\delta_{\Phi} \alpha =  - \frac{2e^2}{E^2} \delta_{\Phi}Re \Bigg( \\ 
\sum_{\alpha\neq\beta}^{}\frac{f_\alpha-f_\beta}{\epsilon_\alpha-\epsilon_\beta}
\frac{\epsilon_\alpha-\epsilon_\beta-i\gamma}{\epsilon_\alpha-\epsilon_\beta+\omega-i\gamma}
{\left|<\beta|\phi_{T.F}|\alpha>\right|}^2 \\
+\frac{\gamma}{\gamma+i\omega}\sum_{\alpha} \frac{\partial
f_\alpha}{\partial\epsilon_\alpha}{\left|<\alpha|\phi_{T.F}|\alpha>\right|}^2 \Bigg)
\label{dpol}
\end{multline}

\subsection{Quantum correction to the polarisability in the diffusive regime for the different statistical ensemble}
\label{regimes}

In this section, we examine the differences between the canonical
and grand canonical ensembles. The canonical ensemble corresponds
to the situation for which the number of electrons in the system
is fixed. It is in particular the case for electrically isolated
systems. The chemical potential is determined self-consistently by
the condition $N=\sum_{i}f_0(\epsilon_i-\mu)$, and therefore
depends on the flux through the energy levels. On the other hand,
in the grand canonical ensemble, the system can exchange electrons
with the thermodynamic reservoir, which imposes the value of the
chemical potential. A physical realization consists in connecting
the ring to a large metallic pad. At finite temperature, the
occupation of the energy levels is spread over an energy interval
of the order of the temperature. As a result, the sensitivity to
the flux dependence of the chemical potential is suppressed when
$T\gg \Delta$, where $\Delta$ is the mean level spacing. Therefore
the differences between CE and GCE are expected to disappear with increasing
temperature and frequency. We first discuss the polarisability in the GCE.

\subsubsection{Grand canonical ensemble}

\begin{figure}
\centerline{ \epsfxsize 6cm \epsffile{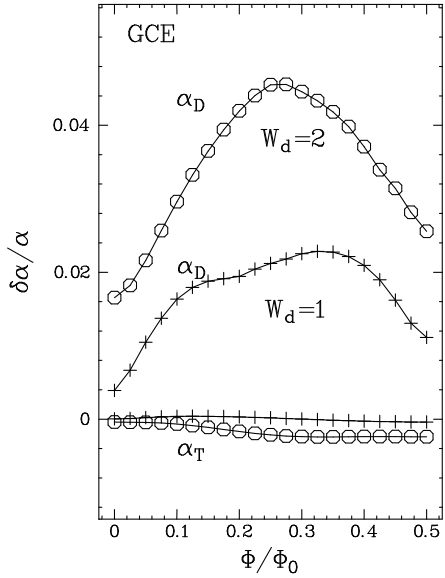}}
\caption{Dynamical ($\omega \gg \gamma)$) and thermodynamical
($\omega \ll \gamma $) polarisability in the grand canonical
ensemble of a disordered ring $80\times 8$, calculated with the
Anderson model for $w=1$ et $w=2$.} \label{xetw1e2}
\end{figure}

Due to the ergodicity property in the diffusive regime, the
disorder average is equivalent to the average over the number of
electrons. This property simplifies the calculation. 
By averaging over the whole spectrum, one can show that $
<\frac{f_\alpha-f_\beta}{\epsilon_\alpha-\epsilon_\beta}>=-\frac{1}{\delta \mu}
$, where $<>\equiv\frac{1}{\Delta \mu}\int_{\mu-\delta \mu/2}^{\mu+\delta \mu/2}$. 
Considering only regimes where $\omega \ll \Delta$ and $\gamma \ll \Delta$ formula( \ref{dpol}) greatly simplifies:
\begin{multline}
\delta_\Phi\alpha = \frac{2e^2}{E_F}\delta_\Phi \left(
\sum_{\alpha\neq\beta} {\left|<\beta
|\phi_{T.F}| \alpha>\right|}^2 \right.\\
\left. +\frac{\gamma}{\gamma+i\omega}\sum_{\alpha}^{}\delta
\left(\epsilon_\alpha-\mu\right){\left|<\alpha|\phi_{T.F}|\alpha>\right|}^2 \right)
\end{multline}
Two frequency regimes can be distinguished, depending wether the
frequency is smaller or larger than the relaxation energy
$\gamma$. In the limit  $\omega\ll\gamma$, which corresponds to a
static electric field , the polarisability is simply equal to the
flux dependence of the trace of the screened potential:
\begin{equation}
\delta_{\Phi}\alpha_s^{GCE}=\frac{2e^2}{E_F}\delta_{\Phi}Tr\left(\phi_{T.F}^{2}\right)
\end{equation}
Since the trace is independent of the basis which is considered,
this quantity is independent of the magnetic flux. Therefore, at
low frequency, there is no mesoscopic correction to the
polarisability in the GCE as shown on figure \ref{xetw1e2}.

 We now turn to the dynamical limit, ie for $\omega
\gg \gamma$. As for the canonical ensemble, we consider a
diffusive ring of radius $R$ and width $W$. The relaxation term
cancels and only remains the term:
\begin{equation}
\delta _{\Phi}\alpha_{D}^{GCE}=\frac{2e^2}{E_F}\delta_{\Phi} \left( \sum_{\alpha >
\beta}^{}{\left|<\beta |\phi_{T.F}| \alpha>\right|}^2 \right) 
\end{equation}
Therefore, one needs to evaluate the matrix element
$\left|\phi_{\alpha,\beta}^{T.F}\right|^2$ .  It is then possible
to use the semiclassical calculation of Mac Millan \cite{macmi}
which yields to  a general expression of the average squares the
matrix elements of the operator $e^{i{\bf q.r}}$ in a diffusive
system which generalizes semi-classical expressions of the matrix
elements of ${\bf X}$ relevant to the unscreened case discussed
above.
\begin{equation}
{\left|<\alpha|e^{i{\bf
q.r}}|\beta>\right|}^2=\frac{\Delta}{\pi\hbar}\frac{Dq^2}{D^2q^4+{\left(\epsilon_\alpha-\epsilon_\beta\right)/\hbar}^2}
\end{equation}

In order to determine the relevant spacial frequencies of the
screened potential, this latter is then decomposed in Fourier
serie: $\phi_{T.F}({\bf r})=\sum_{n=-M/2}^{M/2}A_{\bf n}e^{i{\bf
q^+_n.r}}+e^{i{\bf q^-_n.r}}$, where ${\bf q_n^{+,-}}=\frac{n\pi}{W}{\bf u}_r\pm \frac{2\pi}{L}{\bf
u}_\theta$. At low energy, the main contribution to the matrix
elements of the screened potential  is dominated by the terms with
the smallest wavevector  corresponding to $n=0$,
$q_{min}=\frac{2\pi}{L}$ for which: $\phi_{\alpha,\beta}^2 \approx
\frac{1}{Dq_{min}^2}=\frac{\Delta}{\pi E_c}{\left
(\frac{8R\lambda_s}{3\pi^2W}\right)}^2 $, where $\lambda_s$ is the
screening length, $E_c=\frac{hD}{2\pi R}$  is the Thouless energy and $\Delta$ the mean level spacing
between energy levels.

We can then estimate the flux dependence of the matrix elements $|\phi_{\alpha\beta}|^2$ as was done in section III for the unscreened potential:
\begin{multline}
 \delta_{\phi}<|\phi_{\alpha\beta}|>^2=-\delta_{\phi}<|\phi_{\alpha\alpha}|>^2\\=<|\phi_{\alpha\alpha}|^2>_{GUE}-<|\phi_{\alpha\alpha}|^2>_{GOE}\\ 
=\frac{1}{2}<|\phi_{\alpha\beta}|^2>_{GOE}
\end{multline}

Thus, the mesoscopic correction to the polarisability reads:

\begin{equation}
\frac{\delta\alpha_D^{GC}}{\alpha_0}=\left(\frac{8}{3\pi^3}\right)\frac{\lambda_s}{W}\frac{1}{g}
\label{dpolpol}
\end{equation}
where $\alpha_0$ is the classical value of the polarisability
\footnote{We took the $2d$ limit corresponding to a disk for which
$\alpha_0=\epsilon_0\frac{16}{3}R^3$. For a quasi-1d ring,
$\alpha_0=\frac{\epsilon_0\pi^2R^3}{\ln\left(\frac{R}{W}\right)}$(see
section \ref{polclass}). Only the numerical factor in formula
\ref{dpolpol} is modified if one consider this last value.},
$g=\frac{E_c}{\Delta}$ the dimensionless conductance. This result is in good agreement with the calculation of Blanter and Mirlin  \cite{blant97} using super-symmetry techiques.

The magnetopolarisability increases when a magnetic flux is
applied, corresponding to a {\em positive} magnetopolarisability. A noticeable result is that this effect is
inversely proportional to the conductance and thus increases with
the disorder in the diffusive regime. It is expected however to decrease again with disorder in the localised regime when wave functions and eigen energies become insensitive to the Aharonov Bohm flux.

\begin{figure}
\centerline{ \epsfxsize 6cm \epsffile{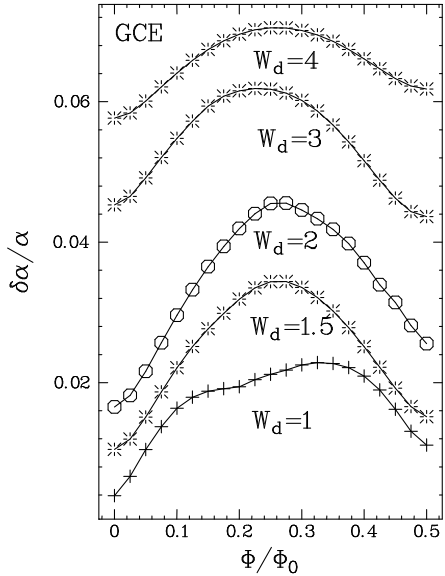}}
\caption{Dynamical polarisability ($\omega \gg \gamma)$) in the
grand canonical ensemble in a disordered ring $80\times 8$,
calculated with the Anderson model for several value of the
disorder: $w=1.$; $1.5$; $2$; $3$; $4$.} \label{xetw}
\end{figure}

This disorder dependence of $\frac{\delta \alpha}{\alpha}$ is
illustrated on figure \ref{xetw}. On the other hand, the
magnetopolarisability decreases with the electron density $n_e$,
since the conductance increases and the screening length decreases
with $n_e$ \footnote{This last point is only true in $3d$: The
screening length decreases with $n_e$ in $3d$ and is independent
of $n_e$ in $2d$.}.

The ring geometry is particularly favorable in order to observe
this effect. The extrapolation of equation(\ref{dpolpol}) leads to
$\delta\alpha/\alpha \sim(\Delta/2E_c)(\lambda_s/a)$ for a two
dimensional sample such as a disk or a square of typical size $a$
and $\delta\alpha/\alpha=(\Delta/2E_c)(\lambda_s/R)^2$ for a
sphere. In a ring etched in a semiconductor heterojunction
GaAs/GaAlAs with the following parameters: $L=8\mu m$,
$\lambda_s=400$\AA, $M=10$, $E_c=7 \Delta$, we obtain $\delta
\alpha/\alpha \sim 3 \times 10^{-3}$.

\subsubsection{Canonical ensemble}

From relation \ref{dpol}, we deduce that at zero temperature and
zero frequency the quantum correction to the polarisability reads
in this ensemble:
\begin{equation}
\delta_{\phi} \alpha = \frac{2e^2}{E^2} \delta_{\phi}\left(\sum_{\alpha =
1}^N\sum_{\beta=N+1}^{N_t}\frac{ {\left|<\alpha|\phi_{T.F}({\bf
r})|\beta>\right|}^2}{\epsilon_\beta-\epsilon_\alpha}\right)
\label{polce}
\end{equation}
Where $N_t$ is the total number of states. Note that this
expression is very similar to the Gorkov Eliashberg one  where
matrix elements of the X operator are replaced by the matrix
elements of the screened potential.  Just as previously discussed in the absence of screening, the flux dependence of the
energy denominators compensates exactly the one of the matrix
element. However this compensation does not exist
anymore, as it has been recently pointed out by Blanter {\it et al}.
\cite{blanter01} when the frequency is of the order of the level
spacing due to the disapearance of the contribution coming from the energy denominators. A similar effect is obtained when increasing the temperature
as shown precisely in \cite{reul94}.  
\subsection{Effect of interactions in the canonical ensemble}

We have shown that in the canonical ensemble in the diffusive
regime, there is no mesoscopic correction to the polarisability.
However, we have neglected in this calculation electron-electron
interactions.

In order to investigate a possible effect of electron-electron
interaction a first and simple approach consists in taking the
interaction potential as a perturbation. Rather than a Coulomb
potential, we use an on-site interaction $U\delta({\bf r-r'})$.
This approximation is justified for high electronic density, since
the interaction between two electrons is then strongly screened by
other electrons. In first order in $U$, the variation in total
energy can be expressed in terms of the local electronic density $n({\bf r})= \sum_{\alpha =1}^N|<\alpha|{\bf r}>|^2$ \cite{rami95}:

\begin{equation}
\delta E_{tot}=U\int n({\bf r})^2dr
\end{equation}

The correction to the canonical polarisability is obtained then by
the formula:

\begin{equation}
\delta \alpha_{int}=-\frac{\partial^2 (\delta
E_{tot})}{\partial^2E}
\end{equation}

\begin{figure}
\centerline{ \epsfxsize 8cm \epsffile{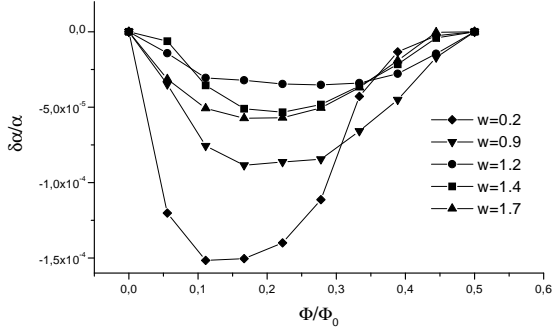}}
\caption{Correction to the canonical polarisability by
interactions in a ring of length $L=60$ and $M=4$ channel, for
several value of the disorder:$w=0.9$; $1.2$; $1.4$; $1.7$.}
\label{deltacop}
\end{figure}

On figure \ref{deltacop} numerical calculations  of this quantity 
$\delta \alpha_{int}/\alpha$  is shown for a ring  for various values of disorder. 

Interactions give rise to a negative magnetopolarisability. The
effect is more important in the ballistic regime than in the
diffusive regime, in which it does not seem to depend  on 
disorder. It gives rise to a magnetopolarisability which is quite
enhanced compared to its value in the absence of interactions.
Nevertheless, the order of magnitude of this effect remains small
compared to the dynamical polarisability.

\section{Comparison with the experiment}

The magnetopolarisability of two dimensional rings has been
recently measured by Deblock et al  \cite{PRL_deblock00}. The
sample consists in an array of rings fabricated by electronic
lithography in an heterojunction GaAs/GaAlAs. The authors use a
resonant technique in which the rings are coupled to the
capacitive part of a high frequency superconducting resonator. 

The polarisability exhibits oscillations as a function of the
flux, with a periodicity corresponding to $\phi_0/2$ through a
ring. The order of magnitude as well as the sign of the effect is
in good agreement with our grand canonical results. In addition,
it decreases as a function of electronic density, according to the
$1/g$ dependence of formula \ref{dpolpol}.

Taking into account the fact that the experiments where done at a
frequency which is of the order of 1/3 of the level spacing, where
differences between canonical and grand canonical results are
strongly reduced, this  result is in good agreement with
theoretical predictions \cite{blanter01}.
\section{conclusion}

We have calculated the average polarisability of
mesoscopic rings and squares. In the grand canonical ensemble, we
found that there exists a positive magnetopolarisability for
frequencies larger than the typical broadening of the energy
levels $\gamma$. The relative effect $\delta \alpha/\alpha$ scales
like $1/g$, where $g$ is the dimensionless conductance. In the
canonical ensemble at zero frequency and zero temperature, the
magnetopolarisability cancels in the diffusive, whereas in the
ballistic regime a small negative effect is found. Differences
between canonical and grand canonical ensemble disappear at
frequencies or temperatures larger than the level spacing. Our
results are in good agreement with recent experiments
\cite{PRL_deblock00}.

\subsection{Summary of the results}

In the following table, are summarized the results obtained for
$\frac{\delta \alpha}{\alpha}$ in the ballistic and diffusive
regime, for the canonical and grand canonical ensemble.

\begin{center}
\begin{tabular}{||c|c|c||}
\hline
\multicolumn{3}{||c||}{Ballistic regime}\\
\hline
  & $ \omega \ll \gamma$ & $\Delta \gg \omega \gg \gamma$\\
\hline
C & $\frac{-1}{4N^2}$ &$\frac{-1}{4N^2}$ \\
\hline
GC & $0$ & $0$\\
\hline
\multicolumn{3}{||c||}{Diffusive regime}\\
\hline
  & $\omega \ll \gamma$ & $\Delta \gg \omega \gg \gamma$\\
\hline
C & $0$ & $0$\\
\hline
GC & $0$ & $\left(\frac{8}{3\pi^3}\right)\frac{\lambda_F}{W}\frac{1}{g}$\\
\hline
\end{tabular}
\end{center}

\appendix
\section{Details on the calculation of the mesoscopic correction to
the polarisability } \label{calculefetov}

In this appendix, we derive in more details the expression of the
mesoscopic correction to the polarisability. In the following, we
will note $\Xi=\Xi_{TF}+\delta \Xi$ and $\xi=\xi_{TF}+\delta \xi$, the response to the external and internal
field respectively, defined in the matrix form as:
\begin{equation}
\delta \rho=\chi \phi=\Xi \phi_{ext}
\end{equation}
The relation \ref{relself} which relate the induced charge density
to the response function can be written in the matrix form:
\begin{equation}
\delta \rho=\chi(1-U\chi)^{-1}\phi_{ext}
\end{equation}
The effective potential is also related to the response function
by \ref{relmat}. The polarisability in the Thomas-Fermi
approximation reads:

\begin{equation}
\begin{array}{lll}
\alpha_{TF}=\frac{1}{E}Tr( x\chi_{TF}\phi)=\frac{1}{E}Tr( \chi_{T.F}(1-U\chi_{T.F})^{-1}\phi_{ext}) \\ \displaystyle =\frac{1}{E}Tr(x\Xi_{cl}\phi_{ext})
\end{array}
\end{equation}

with: $\Xi_{cl}=\chi_{T.F}(1-U\chi_{T.F})^{-1}$.
The next step is to find the quantum correction $\delta \Xi$ to
the screened response function. We set by definition:
\begin{equation}
\Xi=\chi(1-U\chi)^{-1}=\Xi_{cl}+\delta \Xi
\end{equation}
Assuming that quantum corrections are small compared to the
Thomas-Fermi value, second order terms can be neglected:

\[
(1-U\chi_{TF})\Xi-U\delta \chi \Xi_{cl}=\chi_{TF}+\delta\chi
\]

\[
\delta\Xi={\delta \chi}(1-U\chi)^{-1}(1+U\Xi_{cl})
\]
Using the fact that $1+U\Xi_{cl}=\chi_{TF}^{-1}\Xi_{cl}$, we show
that the quantum correction to the response function can be
expressed:

\begin{equation}
\delta \Xi=\Xi_{cl}\chi_{TF}^{-1}\delta\chi \chi_{TF}^{-1}\Xi_{cl}
\end{equation}
The quantum correction to the polarisability :\\
$\delta \alpha=tr (x \delta \Xi \phi_{ext})$
can then be written:
\[
\delta \alpha=tr (\phi_{ TF} \delta \chi \phi_{TF})
\]
where we have used that $ \chi_{TF}^{-1}\Xi_{cl}\phi_{ext}=\phi_{TF}$
\section{Discussion of the linear response approximation}

In order to estimate the validity of the linear response, we have
calculated the total energy of a ring as a function of the
electric field (screening is not here considered). It can be expanded as:

\[
U\left(E\right)=U_0-d_0E-\alpha E^2+...
\]

\begin{figure}
\centerline{ \epsfxsize 6cm \epsffile{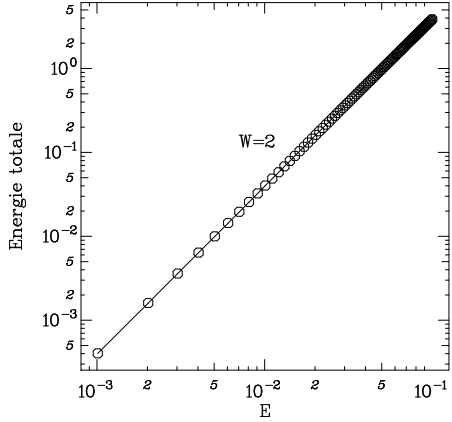}}
\caption{Total energy of the electric field calculated in the
Anderson model in a ring $40\times 5$, and for a disorder $w=2$.}
\label{enerelec}
\end{figure}

The coefficient $d_0$ correspond to the spontaneous dipolar moment
of the ring, resulting from fluctuations of the charge in the
presence of disorder. The ensemble average of $d_0$ is zero. The
energy average varies like $E^2$ at small field. As it
is shown on figure \ref{enerelec} , a deviation from the linear
behaviour is hardly observed before a critical field $E_{max}$ such
that:

\begin{equation}
e E_{max} R \sim E_F
\end{equation}
where R is the radius of the ring.

This critical value is less restrictive than the criteria
\cite{gorko65} and justifies the use of the linear response up to
very high fields. For instance, in a 2D electron gas, obtained in
a semiconductor heterojunction GaAs/GaAlAs for which $E_F \sim
30K$, our criteria allows to use linear response up to fields
such that $E R \sim 2mV$.

\end{document}